# Effective Email Spam Detection System using Extreme Gradient Boosting

Ismail B. Mustapha[a], Shafaatunnur Hasan[a], Sunday O. Olatunji[b], Siti Mariyam Shamsuddin[a], Afolabi Kazeem

[a] Computer Science Department, School of Computing, Universiti Teknologi Malaysia, Skudai 81310 Johor, Malaysia

[b] Computer Science Department, College of Computer Science and Information Technology, Imam Abdulrahman Bin Faisal University, Dammam, Saudi Arabia

Email addresses: bmismail2@live.utm.my, ismail.b.mustapha@gmail.com (I B. Mustapha); shafaatunnur@utm.my (S Hasan); osunday@uod.edu.sa, oluolatunji.aadam@gmail.com (SO.Olatunji), afolabikazeem128@gmail.com (A kazeem)

**Abstract**

The popularity, cost-effectiveness and ease of information exchange that electronic mails offer to electronic device users has been plagued with the rising number of unsolicited or spam emails. Driven by the need to protect email users from this growing menace, research in spam email filtering/detection systems has being increasingly active in the last decade. However, the adaptive nature of spam emails has often rendered most of these systems ineffective. While several spam detection models have been reported in literature, the reported performance on an out of sample test data shows the room for more improvement. Presented in this research is an improved spam detection model based on Extreme Gradient Boosting (XGBoost) which to the best of our knowledge has received little attention spam email detection problems. Experimental results show that the proposed model outperforms earlier approaches across a wide range of evaluation metrics. A thorough analysis of the model results in comparison to the results of earlier works is also presented.

**Keywords** XGBoost, Email Spam, Machine Learning, Spam detector, Computational intelligence

## 1.0 Introduction

Electronic mail (email) is arguably the most popular and cost-effective means of information exchange amongst individuals with electronic devices. Driven by the unprecedented milestone in technological and computational advancement, email has found extensive use in many areas of human endeavour which includes but not limited to job advertisement and recruitment, health care communications, bank transactional information and inter/intra-organizational correspondence. Despite these positives, email usage has generally been plagued with unsolicited and occasionally fraudulent emails popularly called spam or junk emails. While most spam emails are for commercial purposes, a reasonable portion of them are disguised to maliciously mislead or sometimes defraud the recipient; threatening the essence of email as a reliable messaging tool.

The continued rise in spam emails has inspired mitigative approaches to protect email users by filtering or rejecting them all together. This fact is corroborated by a recent Kaspersky Lab Spam and Phishing report which estimates 56.63% as the average amount of spam in email traffic for the year 2017 [4]. Thus, the need for spam filtering which basically entails isolating spam from non-spam emails using computational tools. An even more important stage in the filtering process is detecting that an email is spam or not because this determines what is done to the email afterwards.

Given the fact that spam emails are composed to imitate legitimate ones, their accurate detection has been an active area of research with marked rise in the use of computational intelligent methods in the last decade [5]. However, spammers have increasingly over the years adapted spam emails to appear like legitimates ones. Consequently, the adaptive nature of spam emails have often deceived some of the most effective spam filters; hence the continued need for more accurate spam detection tools still remains an open area of research [1]. Although, many spam detectors have been reported in literature, the predictive accuracies in the various researches suggests the need for better methods with improved accuracy. Bayesian classifiers are one of the earliest methods for spam email filtering and recommendations have been made to facilitate its viability for practical use [8]. For instance, [9] proposed the duo of Artificial Neural Networks (ANN) and Bayesian classifier-based spam filters with impressive spam detection accuracies. In a more recent study, [7] carried out a comparative analysis of 14 different machine learning algorithms and Rotation Tree algorithm performed best on an out of sample test set. Other algorithms like J48, Bayesian Logistic Regression and Multi-Layer Perceptron were also reported with relatively good performance. Meanwhile, [1] introduced a Support Vectors Machine (SVM)-based spam detector using the same dataset under similar experimental setting and obtained a better accuracy on the testing set compared to previous works on the same dataset. In a similar research but different experimental setup on the same dataset, [6] proposed a spam detector based combination of Two-Step Clustering Algorithm with Logistic Regression Method and only succeeded in getting the an accuracy of 93.03% and 98.37% using all and selected features respectively. In addition to reporting results for 10-fold cross validation as against an out of sample validation set, another major shortcoming of this work is lies in the reliance on accuracy as a measure of model performance despite the inherent imbalance in the data. Other computational methods and their hybrids have also been explored for an improved spam detector using the same benchmark dataset [3, 10-12]. Of notable importance is the work of [3], where a hybrid of negative selection algorithm (NSA) and particle swarm optimization (PSO) was used for spam prediction; yielding an improved performance on the testing set compared with the individual method singly. Chikh and Chikhi [2] also presented an improvement to this approach in a more recent work, where a hybrid of clustered NSA and fruit fly optimization (FFO) were combined for spam email detection. A careful look at the afore-highlighted performances exposes the availability of room for performance improvement of the spam email detection approaches as well as the need for more encompassing evaluation metrics that take into consideration the intrinsic class imbalance in the dataset under comparable experimental setting.

In this work, Extreme Gradient Boosting (XGBoost)-based spam email detection model is proposed and investigated for an improved spam detection accuracy. XGBoost [13] is an efficient and scalable variant of the Gradient Boosting Machine (GBM) [14] that has found widespread application in several Machine learning competitions [15]. Its ease of use and parallelization with impressive predictive accuracy has made it a target tool of many Machine Learning (ML) researchers in the present era of big data. Hence, it has been successfully used to solve many complex problems in different fields impressive performance and unique generalization ability. Some of the notable application areas include intrusion detection [16] bio medicine and drug discovery [17-20], Earth and Environmental Science [21, 22] amongst others. Despite its widespread use, its performance in spam email detection systems and related researches is yet to be explored, to the best of our knowledge. Hence, an open area of research we explore herein using a benchmark dataset that has been used in previous representative works under similar experimental setup. We summaries the contribution of this research as follows

- We propose an improved email spam detection system using Xgboost.
- We compare the performance of the proposed system with previous works on the same data
- We analyze the performance of the proposed system using a wider range of evaluation metrics beyond accuracy which has dominated the previous studies

The remaining part of this manuscript is structured as follows. Section 2.0 describes the materials and methods of this research which features a description of the proposed XGBoost-based model, the dataset, evaluation metrics and the experimental design and model implementation details. The experimental results are discussed in section 3.0 while the research is concluded in section 4.0 and future direction proffered.

## 2.0 Materials and Methods

The main task in every supervised learning problem entails learning the optimal set of parameters, $\theta$, that best explain the true relationship between the explanatory and target components of a given dataset $D = \{X_i, Y_i\}$ for $i = (1, 2...n)$ where $X_i = (x_{i1}...x_{ip})$ represents *i-th* observation of a set of $p$-dimensional explanatory variables and $Y_i = (y_1....y_n)^T$ the corresponding target variable for each of the $n$ observations. The best set of parameters $\theta$ are usually learnt via a training process, which results in a supervised learning model that not only correctly fits or maps the training data $X_i$ to their correct labels $Y_i$ but has the capability to generalize well to unseen data. Depending on the nature of $Y_i$, the supervised learning task could either be a regression problem; where the labels are real numbers or a classification problem; where the labels are in groups or categories. The spam detection problem falls in the classification category, since the aim is to classify emails into one of two categories; spam or non-spam. The general task of supervised learning is the optimization of an objective function, often of the form $obj(\theta) = L(\theta) + \Omega(\theta)$, to measure the fitness of the model on the data. $L(\theta)$ represents the training loss function which measures the predictive power of the model with respect to the training data and $\Omega(\theta)$ is an optional regularization term that guides the model from overfitting. XGBoost as a supervised learning method, is briefly described in what follows.

The proposed XGBoost approach to spam detection is a supervised learning ML which is basically an ensemble of a set of weak Classification and Regression Trees (CART), say K, $\{T_1(x_i, y_i)...T_k(x_i, y_i)\}$. Since a CART is a unique type of decision trees that associates a real score to each of its leaves (target or outcome), the prediction score of each weak tree is summed up and the resulting score evaluated through $K$ additive functions, $f_k$, from the space of all possible CART $F$ as in Equation (1)

$$\hat{y} = \sum_{k=1}^{K} f_k(x_i), f_k \in F \tag{1}$$

The regularized objective function is given by Equation (2) where the first and second terms are the differentiable loss function – a measure of the difference between the predicted $\hat{y}_i$ and target $y_i$, and the regularization term- a measure of the model complexity.

$$Obj(\theta) = \sum_i^n (y_i, \hat{y}_i) + \sum_k^K \Omega(f_k) \tag{2}$$

The regularization term is defined by $\Omega(f) = \gamma T + \frac{1}{2}\lambda \sum_{j-1}^{T} w_j^2$ where the vector of scores on each of the leaves and number of leaves are represented by $w$ and $T$ respectively. The additional constants $\gamma$ and $\lambda$ are included to control the degree of regularization. Other methods employed to prevent overfitting in XGBoost are shrinkage and feature/data subsampling [13].

The additive training employed in XGBoost implies that the prediction $\hat{y}_i$ at step $t$ can be formulated as Equation 3. The resulting objective function of the t-th tree after taking the Tailor's expansion of the loss function and expanding the regularization term is shown in Equation 4

$$\hat{y}_i^{(t)} = \sum_{k=1}^{K} f_k(x_i) = \hat{y}_i^{(t-1)} + f_t(x_i) \qquad (3)$$

$$Obj(\theta)^{(t)} = \sum_{i=1}^{n}[g_i f_t(x_i) + \frac{1}{2}h_i f_t(x_i)^2] + \gamma T + \frac{1}{2}\lambda \sum_{j=1}^{T} w_j^2$$

$$= \sum_{j=1}^{T}[(\sum_{i \in I_j} g_i)w_j + \frac{1}{2}(\sum_{i \in I_j} h_i + \lambda)w_j^2] + \gamma T \qquad (4)$$

where $I_j = \{i \mid q(x_i) = j\}$ is the instance set of leaf $j$. For a tree structure $q(x)$, the optimal leaf weight $w_j^*$ and objective function are obtained through Equations 5 and 6

$$w_j^* = -\frac{G_j}{H_j + \lambda} \qquad (5)$$

$$Obj^* = -\frac{1}{2}\sum_{j=1}^{T}\frac{G_j^2}{H_j + \lambda} + \gamma T \qquad (6)$$

where $G_j = \sum_{i \in I_j} g_i$ and $H_j = \sum_{i \in I_j} h_i$. Hence, each leaf is split into two and the score gain calculated with equation 7

$$Gain = \frac{1}{2}\left[\frac{G_L^2}{H_L + \lambda} + \frac{G_R^2}{H_R + \lambda} + \frac{(G_L + G_R)^2}{H_L + H_R + \lambda}\right] - \gamma \qquad (7)$$

Where the first, second and third terms represents the scores of the left, right and the original leaf respectively while the last term is regularization on the additional leaf. Comprehensive details can be found in [13]. A summary of the XGBoost training is as follows;

    i. For each feature/attribute,

        • Sort the numbers

        • Scan the best splitting point (lowest gain)

    ii. Choose the feature with the best splitting point that optimizes the training objective

    iii. Continue splitting (as in (i) and (ii)) until the specified maximum tree depth is reached

    iv. Assign prediction score to the leaves and prune all negative nodes (nodes with negative gains) in a bottom-up order

    v. Repeat the above steps in an additive manner until the specified number of rounds (trees K) is reached.

## 2.1 Data Description

A benchmark spam dataset, from [23], that has found widespread use in representative researches in spam email detection has been used in this work. It is made up of 57 continuous valued descriptive features and one nominal feature which represents the target variable where 1 and 0 denote spam and non-spam emails respectively. The data consists of 4601 data instances from which 1813 (39.4%) are spam and 2788 (60.6%) are non-spam. Further details on the data can be found in [23].

## 2.2 Evaluation Metrics

To ensure fair comparison with previous works, the following commonly used metrics are used to assess the performance of the proposed method.

$$Accuracy = \frac{(TP+TN)}{(TP+TN+FP+FN)} \tag{8}$$

$$Specificity = \frac{TN}{TN+FP} \tag{9}$$

$$Sensitivity\ (Recall) = \frac{TP}{TP+FN} \tag{10}$$

$$Precision = \frac{TP}{TP+FP} \tag{11}$$

$$F1\ score = 2\frac{Precision*Recall}{Precision+Recall} \tag{12}$$

$$Balanced\ accuracy = \frac{Sensitivity+Specificity}{2} \tag{13}$$

where TP, TN, FP and FN stand for true positive, true negative, false positive and false negative respectively.

## 2.3 Experimental Design and Model Implementation

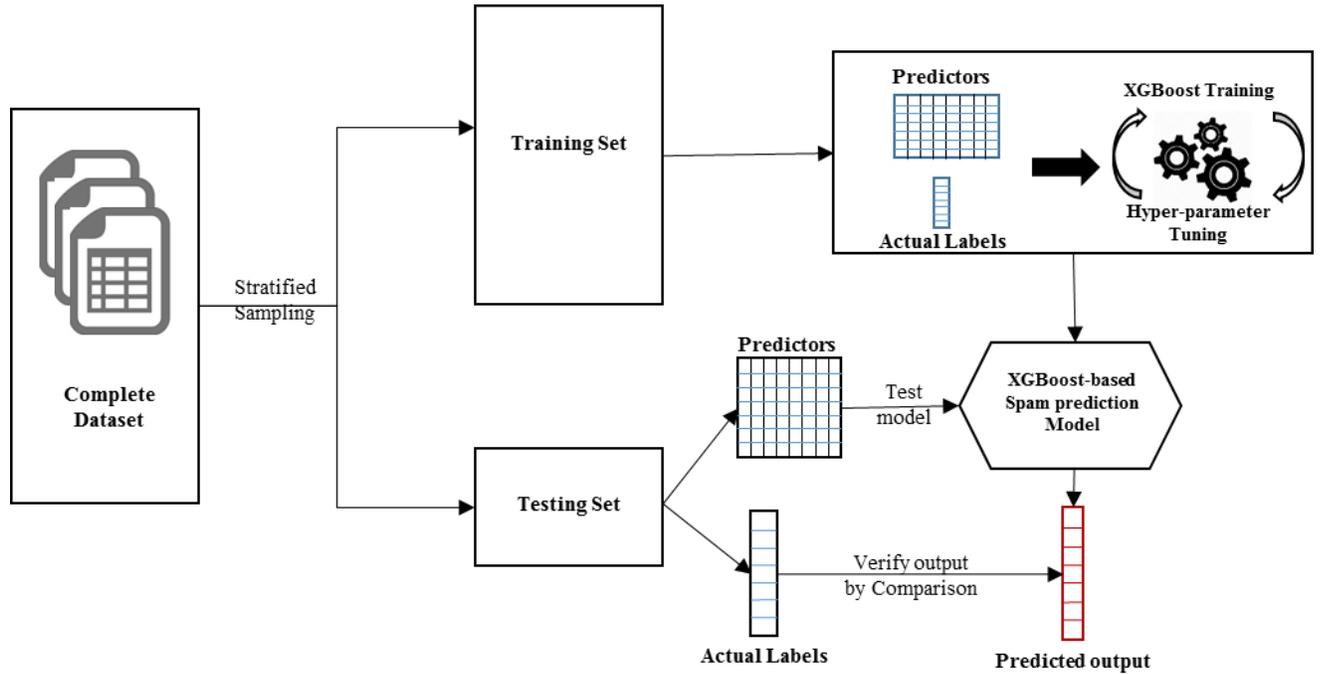

**Figure 1:** Experimental Design for the proposed XGBoost-based Spam Detector

The experimental design and model implementation for the proposed XGBoost-based spam detector is conceptualized in Figure 1. Using a stratified sampling approach, the acquired dataset was split into a training set and an out-of-sample testing set on a ratio 7:3 respectively. While the essence of the 70% training set is to expose the training model to nature and variety of spam and non-spam emails respectively, the 30% testing set is to ensure the performance of the proposed model is tested on data samples it has not come across before.

The model implementation was carried out in R programming environment [24]; with the XGBoost package employed in the modelling process [13]. The training set with the actual labels was fed into the model to train using leave-one-out cross validation. This training approach simply trains the model with a subset of the training set and validates on the rest. This approach is important for hyper-parameter optimization as the performance of the optimized model can be validated during training. Thus, the reason why hyper-parameter optimization was simultaneously carried out during model training.

Moreover, identifying the optimal hyper-parameters for a Machine learning model on a given dataset can be challenging and XGBoost is not an exception in this regard. In fact, this task could even be more challenging when building XGBoost models because of the wide range of tunable hyper-parameters that are available. Therefore, a systematic approach of exhaustive search through specified subset of selected hyper-parameter was carried out using grid search for the spam detection model. The procedure for the grid search is algorithmically highlighted in the steps that follows;

1. *Define specified subset/range of selected hyper-parameter (eta, gamma, minimum child weight etc.)*
2. *Sequentially search through the different combination of hyper-parameters one at a time*
    a. *Train using the next set of hyper-parameters in the sequence step 2 and record performance*
    b. *Return to step 2 if the list of hyper-parameter combinations is unexhausted*
3. *Identify the best hyper-parameter combination and building model*

Given the wide range of tunable hyper-parameters in XGBoost and the time required to optimize them all, we have restricted the grid search to only a few of the hyper-parameters for boosting tree performance as in previous research with Xgboost [17]. The proposed spam detector performed best when the combination of the values of eta, gamma, maximum depth and column sample had the values presented in Table 1. Eta is the step size shrinkage meant to control the learning rate and over-fitting through scaling each tree contribution, gamma is the minimum loss reduction required to make a split, maximum depth is the maximum depth of a child and column sample is the subsample ratio of columns used when constructing each tree. The number of rounds (nrounds) which indicates the number of trees to grow was also set to 200 with early stopping after 10 rounds of no improvement to the training error. This regularization is to ensure the training data is not overfitted. Other three booster parameters like minimum sum of instance weight needed in a child and the proportion of data instance to use when building trees were set to 1.

**Table 1: Optimal combination of parameters for the proposed XGBoost-based Spam Detector**

| | |
|---|---|
| Eta | 0.4 |
| gamma | 0.2 |
| maximum depth | 24 |
| column sample | 0.75 |
| Number of rounds (early stopping after 10) | 200 |

Based on the obtained hyper-parameters, the final spam prediction model was built and tested with the out-of-sample set as illustrated in Figure 1. Experimental results are discussed in what follows in detail.

### 3.0 Experimental results and discussion

The performance of the proposed XGBoost spam detector on the test set is presented and compared with representative researches in spam detection that have evaluated their model using the same dataset. Table 2 shows the confusion matrix from which the evaluation metrics of the proposed XGBoost spam detector model was computed.

Table 2: (a) Confusion matrix for the training set; (b) confusion matrix for the testing set (Note: in the confusion matrix below, 1 stands for spam email while 0 stands for not spam)

| (a) | 0 | 1 |
|---|---|---|
| **0** | 1946 | 1 |
| **1** | 8 | 1266 |

| (b) | 0 | 1 |
|---|---|---|
| **0** | 817 | 24 |
| **1** | 19 | 520 |

Table 2 shows the confusion matrix for the training (a) and test (b) sets whose corresponding TP, TN, FP and FN are 1266, 1946, 1 and 8 and 520, 817, 24 and 19 respectively. From these values, the percentage of each evaluation metric as given in Section 2.2 is presented in Table 3 for the test sets. While the results unsurprisingly reveal an almost perfect classification performance on the training data by proposed XGBoost spam detector model, its impressive performance on the test data shows the generalization capability of the model to an unseen data. Unlike in some machine learning tasks, where good training performance does not translate to similar performance on an unseen test set, the proposed model sidesteps this problem; thanks to the regularization of the training and optimization of the hyper-parameters. This training approach allowed the proposed model to learn the data without overfitting. Given the imbalanced in class distribution of the dataset, receiver operating characteristic (ROC-AUC) and Precision-Recall curve (PR-AUC) (Figure 2) are also presented. The importance of the PR-AUC cannot be overemphasized as it has been reported to be more informative when it comes to imbalanced data [27].

Table 3: Test Results in Percentage

| Data | Sensitivity/ Recall | Specificity | Precision | F1-Score | Balanced Accuracy | Accuracy | ROC-AUC | PR-AUC |
|---|---|---|---|---|---|---|---|---|
| **Test** | 95.59 | 97.73 | 96.47 | 96.03 | 96.66 | 96.88 | 99.08 | 97.69 |

In addition, the performance of the proposed XGBoost spam detector model is compared with recent and previous models that have been evaluated on the same dataset. As shown in Table 4, the proposed model with testing accuracy of 96.88% clearly outperforms the best reported accuracy in literature from earlier works by as much as 2.46%; making 94.06% the best reported accuracy for an SVM-based spam detection model in [1]. Other notable works relating to spam detection with similar experimental setup and dataset that the proposed approach outperforms include schemes based on hybridized two-step clustering and logistic regression algorithm[6], hybridized clustered negative selection algorithm–fire fly optimization (CNSA-FFO)[2] and other variants of hybridized NSA, particle swarm optimization (PSO) Rotation Forest, MLP and J48[7]-based classifiers. It is however, worth mentioning that a performance accuracy of 98.41% was reported on the same dataset in [6]. We do not find this result to be comparable to ours due to the disparity in experimental setup.

Figure 2: (a)ROC and (b) PR Curves

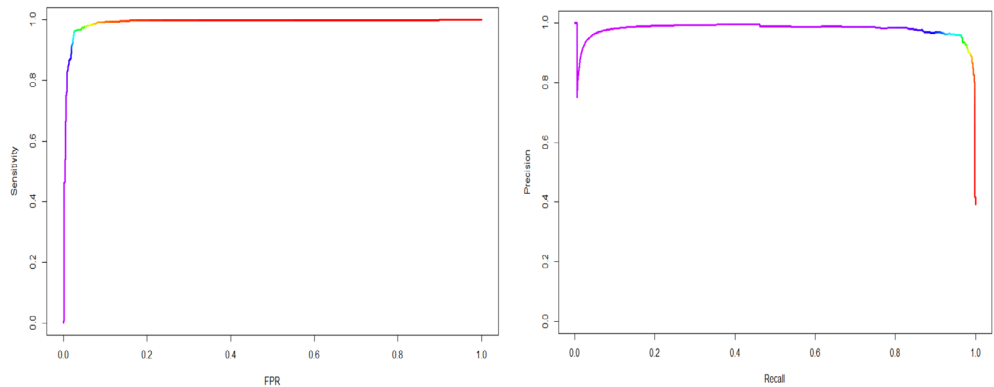

This supports the success that XGBoost has been reported to enjoy in other fields[17, 22, 25, 26]. Thus, given the performance of the proposed XGBoost spam detection model as an outstanding discriminator of spam and non-spam email, it would be a vital addition to the already wide array of computational intelligent approaches for spam email detection.

Table 4: Comparison of testing performance of the proposed XGBoost- based spam detector and other earlier published classifiers on the same dataset

| Classifier | Accuracy | Sensitivity/ Recall | Specificity | Precision | F1-Score | ROC-AUC |
|---|---|---|---|---|---|---|
| Proposed XGBoost | 96.88 | 95.59 | 97.73 | 96.47 | 96.03 | 99.08 |
| SVM[1] | 94.06 | 93.87 | 94.06 | - | - | - |
| CNSA-FFO[2] | 93.88 | 87.28 | 97.31 | - | - | - |
| NSA–PSO[3] | 91.22 | 65.99 | 93.43 | - | - | - |
| PSO[3] | 81.32 | 60.48 | 94.86 | - | - | - |
| NSA[3] | 68.86 | 22.24 | 99.16 | - | - | - |
| LR-Two-step[6] | 93.03 | - | - | - | - | - |
| LR[6] | 90.85 | - | - | - | - | - |
| Rotation Forest[7] | 93.50 | 93.50 | - | 93.50 | 93.50 | 97.60 |
| J48[7] | 91.20 | 91.20 | - | 91.20 | 91.10 | 93.70 |
| Bayesian LR[7] | 93.00 | 93.00 | - | 93.00 | 93.00 | 92.70 |
| MLP[7] | 92.30 | 92.30 | - | 92.30 | 92.30 | 97.30 |

## 3.1 On the Impact of Imbalance on Model Learning

Learning from imbalanced data have been thoroughly studied and still remains an active area of machine learning [27]; with some studies focusing on the effect of skewed class distribution and other contributing factors on the performance of supervised machine learning problems [28, 29].

In this study, we examine the impact of imbalance in data distribution on the performance of the proposed XGBoost spam detection model. Several data level approaches to imbalance data resolution which include Random over/under-sampling, SMOTE, Borderline SMOTE, ADASYN, SMOTENN, Tomek links and SMOTE-Tomek links were applied to the training set, trained under the same experimental setup and tested on an imbalanced out of sample test set. We find that the performance of the model does not increase across all the various data-level approaches. Hence, we have omitted the results given that it does not improve it does not what has been reported above.

## 4.0 Conclusion and Future Directions

XGBoost, as a detector of spam emails, has been proposed in this work using standard computational intelligence approach on a benchmark dataset and experimental results have been shown to be an invaluable classifier which with optimized hyper-parameters is less susceptible to overfitting. The XGBoost spam detection model has been built with a training set and tested with an out-of-sample test set and its remarkable performance of 96.88% accuracy outperforms the best reported accuracy by other spam detection schemes on the same dataset and similar experimental setup; reaching a performance improvement of 2.82% over an SVM-based spam detector reported to be the highest so far [1].

A notable limitation of this work is that no explicit feature selection or extraction was carried out on the dataset in the experimental process. This is mainly to have a balanced assessment of the performance of XGBoost as a standalone spam detector. Given the fact that the speed and performance of classification in other domains often increased by reducing the feature space of the dataset, a viable future direction is to explore the possible performance improvement that can be achieved through feature reduction/extraction methods with XGBoost.

In addition, another important future direction of this work is to evaluate XGBoost on more spam datasets and compare the performance with popular machine learning algorithms.

**Compliance with ethical standards**

**Conflict of Interest:** The authors declare that they have no conflict of interest.